# Guiding superconducting vortices by magnetic domain walls

V. Vlasko-Vlasov, U. Welp, G. Karapetrov, V. Novosad, D. Rosenmann, M. Iavarone, A. Belkin[1], W. –K. Kwok

Materials Science Division, Argonne National Laboratory, Argonne, IL 60439
[1]MSD ANL and Physics Division, Illinois Institute of Technology, Chicago, IL 60616

We demonstrate a unique prospect for inducing anisotropic vortex pinning and manipulating the directional motion of vortices using the stripe domain patterns of a uniaxial magnetic film in a the superconducting/ferromagnetic hybrid. Our observations can be described by a model, which considers interactions between magnetic charges of vortices and surface magnetic charges of domains resulting in the enhanced pinning of vortices on domain walls.

Superconducting/Ferromagnetic (SC/FM) hybrids offer a variety of exciting new phenomena, which have been extensively discussed in recent years [1- 3]. They are defined by *short range* and *long range* interactions between the competing SC and FM components responsible for the *paramagnetic* or *proximity* effects (due to FM exchange) and *orbital* coupling (due to magnetic stray field), respectively. In general, the phenomena of superconductivity and ferromagnetism are considered incompatible because singlet superconducting Cooper pairs carry opposite spins while in the ferromagnets all the spins are aligned. However, they can coexist in the bulk under a short-range proximity effect if they form fine scale inhomogeneous structures. Superconductivity could survive under conditions of a weak ferromagnetic exchange interaction by forming a spatially oscillating order parameter , the Fulde-Farrell- Larkin-Ovchinnikov (FFLO) structure [4,5]. In turn, the ferromagnetic component could form oppositely magnetized domains at lengthscales way below the SC penetration depth (cryptomagnetic structure [6-8,3]), which would support "antiferromagnetic" Cooper pairs. Neither FFLO, nor cryptomagnetic structure were convincingly confirmed so far for bulk materials. However, proximity phenomena such as the appearance of π- and intermediate phase Josephson junctions, have been experimentally demonstrated in FM/SC hybrids (see review [9]). The short range of these effects is defined by the oscillating penetration of the SC pairs into the FM and is usually of the order of the coherence length. These scales are similar to the size of cryptomagnetic domains envisioned in bulk FM/SCs [1]. A special case of the proximity effect should appear in FM/SC *bilayers* due to the FM domain walls where the reduced effective exchange at the wall could lead to a locally enhanced superconducting transition temperature, $T_c$, in the SC layer [10].

The *long range* effects in FM/SC bilayers are associated with interactions of the FM stray fields with the SC screening currents. Due to such interactions, more robust superconductivity just above the FM domain walls is expected when domains are magnetized *perpendicular* to the SC/FM interface and the stray fields at the domain wall are at a minimum (see [11] and refs. there-in). In contrast, domain walls between *in-plane* magnetized domains carry enhanced stray fields and should locally suppress the critical temperature. These effects can introduce exotic H-T phase diagrams depending on the presence or collapse of FM domains, which were recently observed experimentally [12-14]. Also, new effects can appear due to the coupling of the FM domains with SC vortices. The coupling is not trivial and results in properties that do not exist in separate SC or FM layers [2]. It can modify the FM domain structure below $T_c$, cause the generation of spontaneous vortices and vortex antivortex pairs, define specific vortex configurations, and form combined domain structures of coupled domains and vortices with the same polarization. One obvious consequence should be the enhancement of the SC critical current and increase of the FM coercivity due to the mutual pinning of vortices and domain walls. Some indications of these effects have been already reported [15-17] and will be discussed in more detail later. Using above effects one can manipulate vortices by changing the domain structure and vary the domain structure by re-arranging vortices.

In this work we show how manipulations with FM domains can be used to form a robust potential for directing vortex motion thus yielding a strong tunable anisotropy of critical currents in the adjacent superconductor. This opens propects for creating new cryogenic devices with improved magneto-electric response where e.g. strong variations of the conductivity near $T_c$ can be induced by relatively small magnetic fields, changing the FM domains.

The FM/SC bi-layer structure was fabricated by sputtering a 0.8μm FM permalloy film (Py: $Ni_{80}Fe_{20}$) onto the surface of a 20μm thick SC 2H-$NbSe_2$ single crystal with critical temperature, $T_c$=7.2K. Magnetization loops of the Py film in the bilayer were measured with a SQUID magnetometer and its domain structure was imaged using room temperature magnetic force microscope. Direct magneto-optical observations of the vortex motion with respect to well controlled domain structures were conducted using the garnet indicator technique [18] after cooling the sample below $T_c$.

Magnetization loops for the Py film are shown in Fig.1. We determined the magnetization $M_s$ and the uniaxial anisotropy constant $K_u$ along the film normal from the perpendicular, $H_\perp^{sat}$ = 11200 Oe, and parallel, $H_{//}^{sat}$ =200 Oe, saturation fields, using formulas for materials with small perpendicular anisotropy [19]:

$$H_\perp^{sat}=4\pi M_s(1-K_u/2\pi M_s^2), \quad H_{//}^{sat}=2K_u/M_s \qquad (1)$$

The resulting values of $4\pi M_s$=11400 G ($M_s$=907 G), $K_u$=90720 erg/cm$^3$, and quality factor of $Q=K_u/2\pi M_s^2$ =0.0175 are in good agreement with magnetic constants reported for Py films of close compositions [20-22].

The growth induced perpendicular anisotropy of the Py film should result in the stripe domain structure above a critical thickness [22] of $t_c=2\pi(A/K)^{1/2}$ =0.21 μm. Here, we used a frequently cited value of the exchange constant for Py, A=1*10$^{-6}$ erg/cm (e.g. [21]). In our 0.8μm thick film, MFM images of the demagnetized state showed a labyrinth domain structure, which is defined by the kinetics of the domain nucleation and known to provide the minimum magnetostatic energy. By applying a strong enough (H>300 Oe) in-plane field we could align the domain walls in a desired direction. This domain structure was maintained after switching off the magnetic field. Such alignment of stripe domains, defined by the reduction of demagnetizing fields and the polarization of domain walls, is well known for films with perpendicular anisotropy with both high [23] and low Q factors [24]. By changing the orientation of the in-plane field, the direction of the stripes can be precisely controlled. Fig.2 shows MFM images of the domain structure in the Py film obtained after successive applications of 1 kOe in two perpendicular in-plane directions. This stripe domain structure is very stable and the domain walls start moving in fields normal to the film surface above the coercivity field $H_c$~117 Oe. However, the parallel orientation of domains remains practically unchanged until the stripe-collapse field of $H_{sc}$~±400 Oe. This field, which is larger than $H_c$, defines the range of fields where the effects described below are valid. At room temperature, the variation of the domain width was 0.39 to 0.41 μm as observed after several applications of the in-plane field in different directions. This is close to the predicted stripe width [19] d= $(\pi t)^{1/2}[(1+Q)A/K_u]^{1/4}$ = 0.3 μm for the above magnetic constants and film thickness t=0.8μm and confirms our value of $K_u$. Low temperature values of the constants can be obtained using Bloch theorem $M_s=M_s(0)[1-\alpha T^{3/2}]$ and $K_u=K_u(0)[1-\alpha T^{3/2}]^3$ with $\alpha=2.78*10^{-5}$ 1/Deg$^{3/2}$ as estimated from the temperature variation of $M_s(T)$ in Py films [25]. This yields $M_s(0)$= 1054 G, $K_u(0)$= 142260 erg/cm$^3$ , and accounting that A is nearly temperature independent, we find d(T=0)=0.27 μm, which is reduced only slightly from the room temperature value.

Magnetic flux penetration and exit patterns were observed at temperatures below $T_c$ of the SC NbSe$_2$ crystal after cooling the sample with the aligned stripe structure and application of the magnetic field perpendicular to the sample surface. Figs. 3, 4, and 5 illustrate our observations for three orientations of stripe domains prepared by the application and switching off of the in-plane magnetic fields at an angle of ~45° to the sample edge, and also parallel, and perpendicular to the edge, respectively. The figures present successive flux distributions in the zero-field cooled sample at increasing and decreasing perpendicular magnetic fields $H_z$. The flux entry in the pictures is delineated by the magnitude of the bright contrast which corresponds to the normal magnetic induction $B_z$ associated with the local vortex density. Vortices start penetrating in weak spots near the sample edge and spread inside the sample with increasing $H_z$. The same weak spots act as the flux exit points (dark contrast) with decreasing $H_z$.

The remarkable feature of all three FM domain orientations is a distinct preferential vortex motion along the stripes indicated by the arrows in the figures. This behavior does not change after cycling the temperature above and below $T_c$ and is maintained up to reasonably high $H_z$ below the domain-collapse field of $H_{sc}$~400 Oe. The emerging flux patterns are practically the same even when the stripes are aligned by the application and switching off of the in-plane field at $T<T_c$ demonstrating that the pinning effect of in-plane vortices that could be introduced during the alignment of stripe is smaller than the pinning/channeling effect on the vortices due to the FM domains. In normal fields of ~170 Oe flux gradients are hardly observable due to the small intrinsic pinning in NbSe$_2$, but at successive decrease of the field the preferential exit of vortices along the initial stripe direction (see dark regions in Figs.3-5 at $H_z$ reduced to zero) is quite obvious.

The alignment of stripes parallel (Fig.4) and perpendicular (Fig.5) to the edge of the sample caused a noticeable expansion and narrowing of the flux entry pattern, respectively. Obviously, the magnetic domain walls aligned parallel to the edge introduce an additional pinning for vortices moving from the edge, while there is no additional pinning for vortices moving along domain walls in the case of stripes perpendicular to the edge. The ratio of the flux penetration distances observed at the same $H_z$ for these two perpendicular orientations of the stripes gives an estimate of the critical current anisotropy γ introduced by domain walls. A comparison of flux patterns in Fig.4 and 5, e.g. at $H_z$=27.5 Oe, yields γ~3. Such increased currents along the stripe domains showing the dominating role of magnetic pinning in the hybrid are observed in a wide temperature range below $T_c$. The effect is robust and the direction of the easy/hard vortex motion remains unchanged even after cycling the sample to room temperature and cooling it back down. Furthermore, the easy/hard vortex entry pattern remains the same over an extended time period, confirming the stability of the initial field induced orientation of stripe domains in Py .

The effect of domains on the enhancement of pinning in the FM/SC bilayer was considered theoretically [26, 27] and then sought-after experimentally by a number of groups. In a hybrid of SC YBa$_2$Cu$_3$O$_{7-d}$ and FM BaFe$_{17}$O$_{19}$ films separated by a thin yttria-stabilized – zirconia buffer layer a strong upward shift of the irreversibility line, $H_{ir}(T)$, compared to pure

YBCO films was observed [15]. This increase of the irreversibility field near $T_c$ was associated with the magnetic pinning on stripe domains, although the total magnetic moment of the hybrid (and thus the SC current which should increase due to the additional magnetic pinning) was reduced by a factor of ~200 relatively to the pure YBCO film. A more thorough control of the FM domain structure in a bilayer thin film of Pb on CoPt has revealed a 3-fold increase of the low field critical current due to nucleation of bubble domains in the CoPt film [28]. The subsequent expansion of the bubble domains into a labyrinth pattern resulted in a strong suppression of SC currents due to the magnetic domain stray fields. In YBCO/CoPt bilayers a 2-3 times increase of the critical current compared to a single YBCO film was observed at 86K [29]. Remarkably, even at H=1T, when domains are expected to collapse, the enhanced $J_c$ was still visible at 86K (Fig.2 in [29]), although at 75K, $J_c$ decreased at small fields compared to the single YBCO layer. Similarly, a 1.6 times decrease of $J_c$ at 4.7K in the case of stripe domains in a Nb/Cu/SrRuO$_3$ hybrid was found when compared to uniformly polarized FM SrRuO$_3$ [30]. The suppression of $J_c$ was linked to the stray fields of the magnetic domains. In a YBCO film covered with an in-plane anisotropic FM Pr$_{0.67}$Sr$_{0.33}$MnO$_3$ layer [31], a ~1.5-fold enhancement of the critical current at T>50K as compared to the pure YBCO film was estimated from the hysteresis magnetization loops measured in perpendicular fields. This enhancement, which increased to ~1.7-fold at 3 kOe, and even to > 2-fold at temperatures near 70K (Fig.7 of [31]), was associated with vortex pinning due to the underlying magnetic domain walls. However, it is unclear why boundaries of the in-plane magnetized domains improve their pinning efficiency with increasing perpendicular fields. A 2.5 times increase of pinning (compared to the saturated FM state) in fields below ~50 Oe and at ~0.9$T_c$ was reported for a Nb/CoPt bilayer when the residual dendrite-shaped domains were formed at the final stage of the magnetic reversal of CoPt [17]. Recently, the magnetic flux penetration was imaged magneto-optically in YBCO/La$_{0.67}$Sr$_{0.33}$MnO$_3$ bilayers, where inhomogeneous magnetic patterns were induced in the FM LSMO by the sample's twin structure [32]. In this case, the enhanced vortex entry along twin boundaries with stray fields of the same polarity as the applied magnetic field and the delayed vortex penetration at boundaries of opposite magnetic polarity were observed.

Our data directly show that FM domains introduce additional pinning, which results in a ~3-fold enhancement of the critical current along the stripe domains. In addition to observations of the preferred vortex motion along the stripes, we also demonstrate a new prospect for controlling this process by reorienting the stripe pattern in a desired direction. We relate the observed phenomenon to domain wall pinning, which induces a critical current anisotropy with enhanced $J_c$ along the walls. Such an enhancement of pinning due to FM domains was first suggested in [26] and was later calculated in a number of theoretical works for different parameters of the FM and SC [27, 33-37]. A clear physical picture relevant to our 'thick' FM/SC bi-layer with perpendicular anisotropy of the FM film is presented in [35]. It considers interactions of vortices with the domain structure modified by the presence of a superconductor. In our case, the Py (0.8 μm) and NbSe$_2$ (20 μm) thicknesses are essentially larger than any FM and SC length scales: penetration depth of NbSe$_2$ $\lambda_{ab}$~130nm [38], coherence length $\xi_{ab}$=7.9nm and $\xi_c$= 2.4nm [39], domain width ~0.4 μm, and estimated domain wall thickness $2\pi(A/K)^{1/2}$ =0.21 μm, and exchange length $(A/2\pi M^2)^{1/2}$=4.4nm. At H=0, the spontaneous nucleation of vortices is expected for high enough magnetization in domains producing strong enough stray fields outside the FM. Straight vortices can appear in the center of the domains and vortex loops can form around the domain walls. In high-Q materials, the stray fields are described by the periodic distribution of magnetic charges $\pm4\pi M_s$ in neighboring domains, arising from the discontinuity of the magnetization at the surface as shown in Fig.6a. Below $T_c$, when the magnetic fields of the domains are screened by the SC substrate (Fig.6b), the stray fields will double in the FM near the SC/FM interface (in the approximation of zero SC penetration depth $\lambda$) but occupy only half of their previous volume, so that their total energy($\sim H^2 V$) also doubles. On the other surface it remains the same (if d<h, interactions between top and bottom surface charges are negligible). Hence the total demagnetization energy increases 1.5 times, which can result in a $(1.5)^{1/2}$ shrinkage of the domain width [34]. A finite $\lambda$ will give a slightly smaller shrinkage value [40]. In reality, one should expect the effect to be smaller due to the coercivity of the domain walls and the partial freezing of the stray fields in the SC when cooled through $T_c$.

It was shown in [41] that the field of a vortex at the SC surface can be approximated by the field of a magnetic monopole with a charge of two flux quanta, $2\Phi_0$, at a distance $\sim\lambda$ below the surface. In the approximation of negligible $\lambda$, the interaction of the FM with vortices can be reduced to the coupling of the FM magnetic charges to the vortex magnetic charge at the FM/SC interface, which results in the interaction energy [35]:

$$U_{VM} = \pm \frac{2\Phi_0 M_s}{\pi} * f(x)$$

where

$$f(x) = \int_0^x \ln(\tan(\pi x/2d)) dx$$

Note that f(x) varies smoothly inside the stripe domains (Fig.6d) while the surface magnetic charge due to domains follows a step function. Both f(x) and its derivative (Fig.6e), which presents the in-plane field $H_x(0)$, increase towards the domain wall and should be cut off at a distance $x\sim 0.5\delta$ (or $\lambda$, if $\lambda > \delta$). An estimate of the maximum pinning force for a $\sim 1$ μm high-Q FM film with $K\sim 10^6$ erg/cm$^3$ and $M_s\sim 10^3$ G where the domain wall width is $\delta\sim 0.6*10^{-5}$cm, and the domain width is $d\sim 1.5*10^{-5}$cm, yields $F_p\sim (2\Phi_0 M_s/\pi)*|\ln(\tan(0.1\pi))|\sim 2.2\Phi_0 M_s/\pi$. For a straight vortex of length $l_v$, this gives a critical current density $J_c\sim 0.7 M_s/l_v$. For 1 μm thick superconducting film this yields $J_c\sim 0.7*10^5$ A/cm$^2$.

In the case of Py, which has a small anisotropy, the domains are not homogeneously magnetized perpendicular to the surface but form a twisted structure schematically shown in Fig.6c (for detailed micromagnetic simulations see e.g. [21-22]). Magnetic charges are not only residing at the surface but are distributed within some layer near the surface. The resulting stray fields are weaker than in the high-Q materials and have a smoother profile. Accepting the sinusoidal surface magnetic potential and stray field distribution (Fig.6g-f) with amplitude taken from numerical simulations in [21] $H_x(0)\sim 0.316 M_s$ we can estimate the pinning force due to the interaction of vortex charge and the FM stray fields [35] as $F_p^M\sim 0.3\Phi_0 M_s/4\pi$. Here we neglected doubling of the magnetic charge at the FM/SC interface assuming that the stray fields of the domains are not screened but rather frozen into the SC as the sample is cooled below $T_c$. For straight vortices in our 20 μm sample, this would correspond to $J_c^{DW}\sim 0.87*10^4$ A/cm$^2$. Taking the observed anisotropy of $\sim 3$ for the flux penetration along and across the domain wall yields an intrinsic critical current of $J_c\sim 2.9*10^3$A/cm$^2$. This is close to the value of $J_c\sim 2$-$2.5*10^3$/cm$^2$ reported for NbSe$_2$ crystals at H=0 and T$\sim 4.2$K [42-43]. Thus our results can be reasonably well explained by the model of [35] if we account for the modification of stray fields in a uniaxial ferromagnet with a small Q-factor.

Conclusions

In summary, we presented a new prospect for introducing a pronounced anisotropy in the vortex motion in SC/FM hybrids by aligning the stripe domains in the ferromagnetic layer. The observed critical current anisotropy of ~3 can be improved by using a thinner SC substrate with smaller intrinsic pinning coupled to a FM film with a larger Q-factor. In fact, for very thin SC films ($d<\lambda$) on a thin FM substrate with wide stripe domains and high Q, the critical currents along the stripes are predicted to be at least 100 times larger than $J_c$ perpendicular to the domain walls [36]. Our observations confirm that domain walls form a strong pinning barrier for vortices. They support conclusions of [44, 45] where the pinning of vortices at domain walls was considered as the main reason of the enhanced critical currents and reduced resistivity in FM/SC hybrids at the FM coercivity field.

By reorienting magnetic domains or changing the symmetry of the domain lattice using combinations of DC and AC fields, it is possible to rearrange current patterns in the FM/SC bilayer and manipulate its conductivity. In addition to directing vortex motion, this method may provide means to vary the pinning strength using the rich variety of domain patterns found in FM films with perpendicular anisotropy, such as labyrinth, stripe, or cylindrical domain lattices, which are stable in appropriate ranges of magnetic fields..

Acknowledgements

This work was supported by UChicago Argonne, LLC, Operator of Argonne National Laboratory ("Argonne").  Argonne, a U.S. Department of Energy Office of Science Laboratory, is operated under Contract No. DE-AC02-06CH11357.

Figures

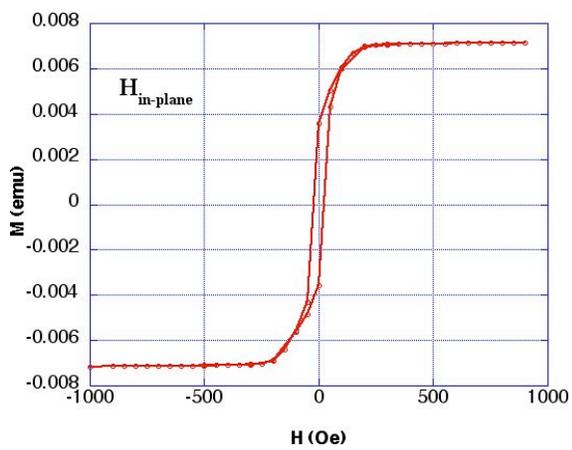 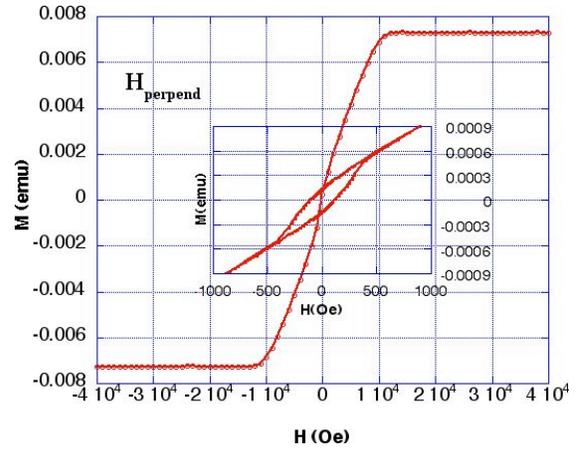

(a)                                                                                  (b)

Fig.1 Magnetization loops of Py film at 290K for in-plane(a) and perpendicular (b) fields. The insert expands the low field region revealing the domain-collapse fields ~±400 Oe.

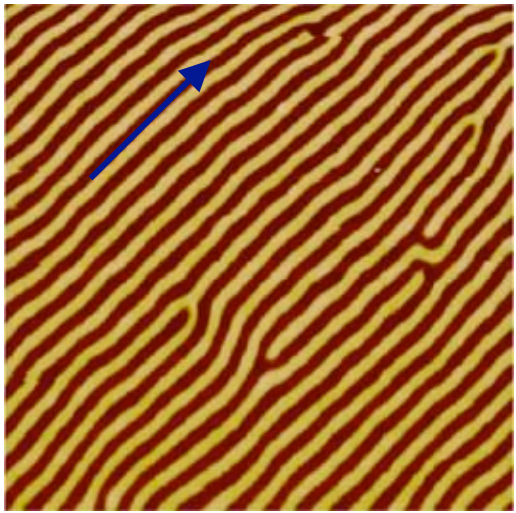 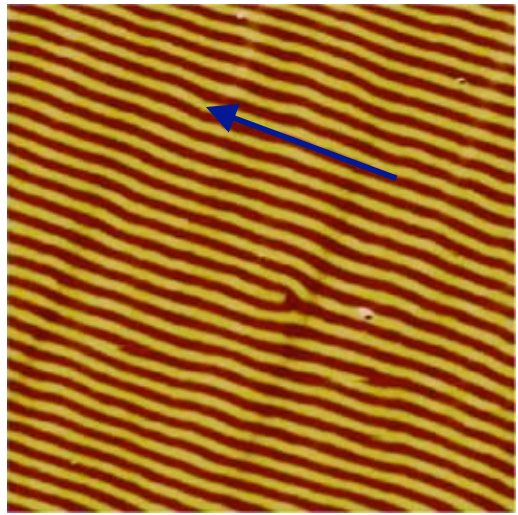

(a)                                                                                  (b)

Fig.2  20x20 µm$^2$ MFM images of stripe domains in Py film after applying and turning off a H=1 kOe in-plane magnetic field along directions indicated by the arrows.

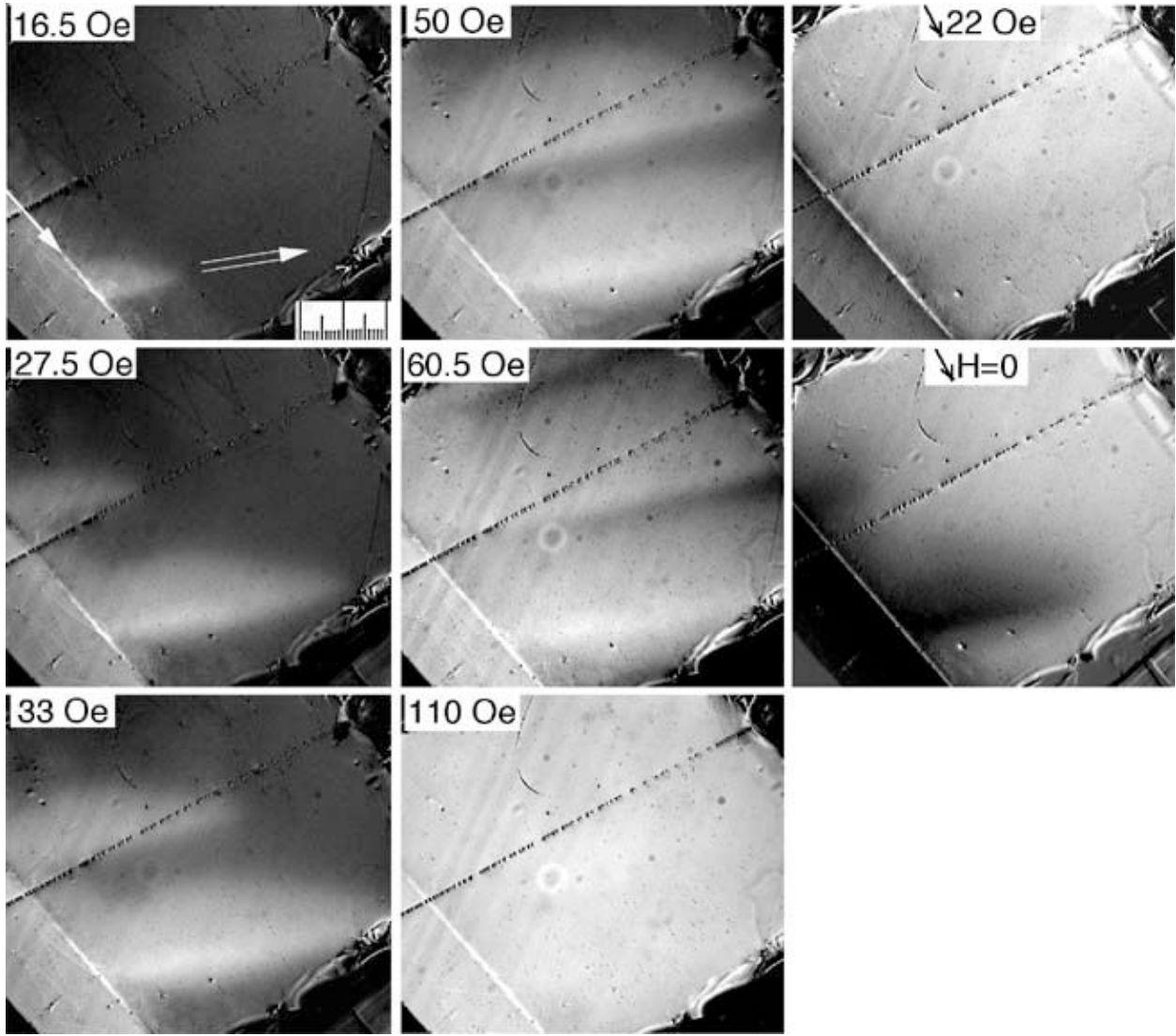

Fig.3 Magneto-optical images of flux entry (with increasing perpendicular field) and exit (two pictures on the right obtained at decreasing field from $H_z^{max}$=250 Oe) at T=4.5K following preparation of the stripe domain structures by turning on and off an in-plane field of H=1 kOe at an angle ~45° with respect to the sample edge at T>$T_c$. The brightness of the magneto-optical contrast corresponds to the vortex density. Values of the perpendicular field normal to the sample surface are shown on the pictures. Double-line arrow shows the preferential flux entry direction related to the direction of the stripe domains in the Py. Single-line arrow marks the sample edge. The diagonal dotted line is a scratch in the indicator film. The scale bar is 200 μm.

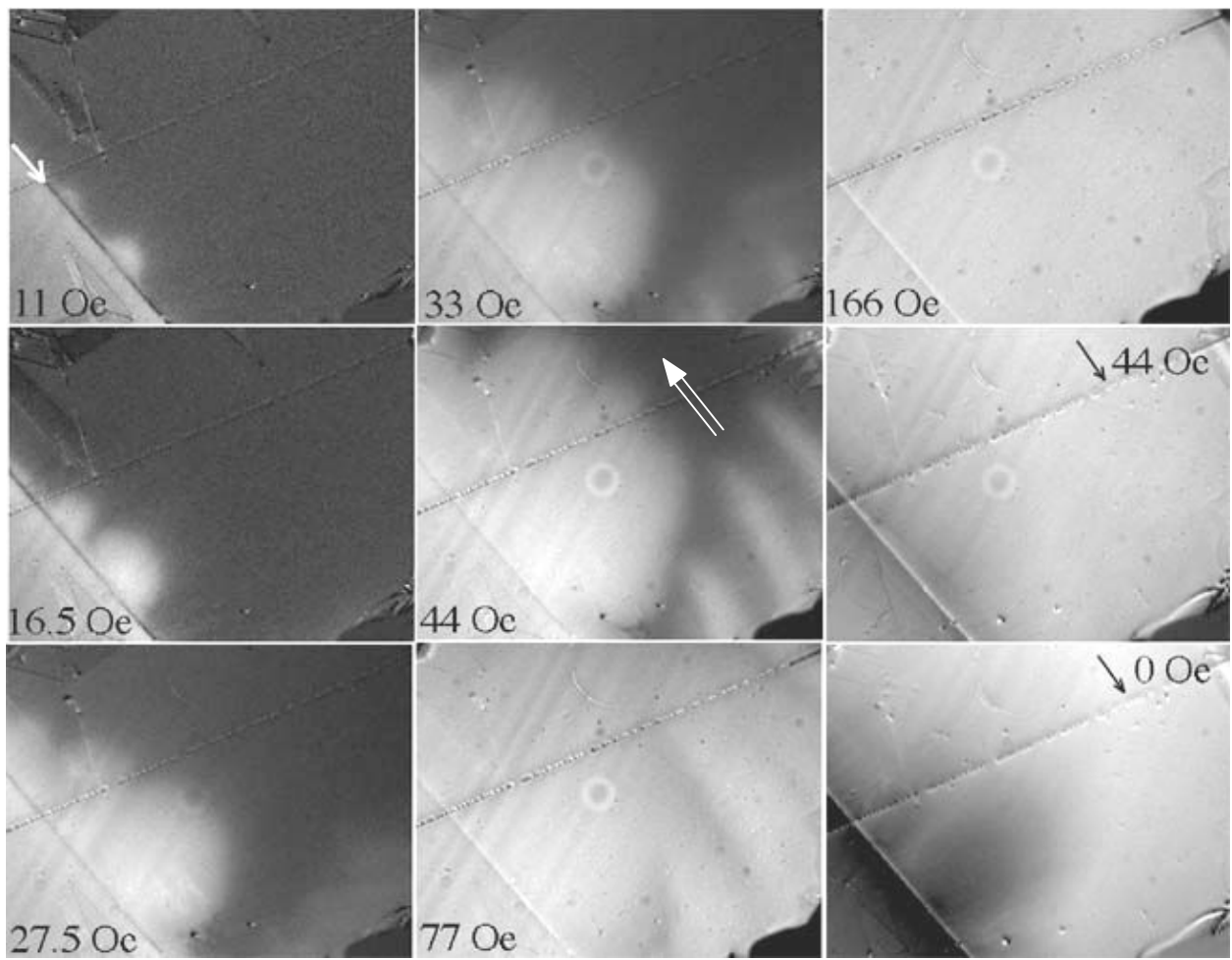

Fig.4 Same as Fig.4 after application and switching off H=1kOe along the sample edge shown by the arrow at $T>T_c$. Two pictures at the right marked with black arrows are taken at decreasing field.

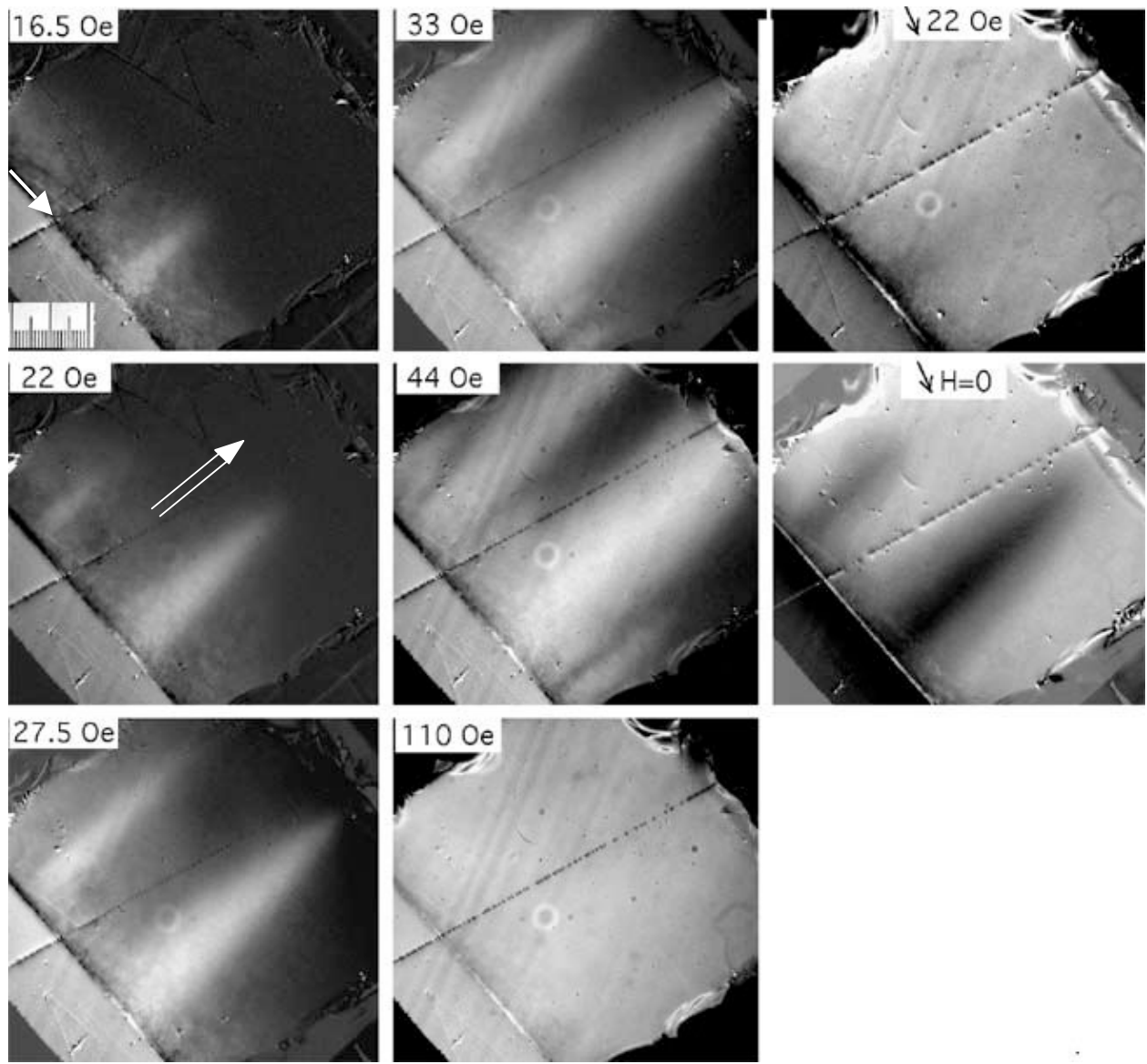

Fig.5 Same as Fig.3-4 after application and switching off the in-plane field H=1kOe perpendicular to the sample edge at $T>T_c$. Two pictures on the right show flux exit at decreasing field.

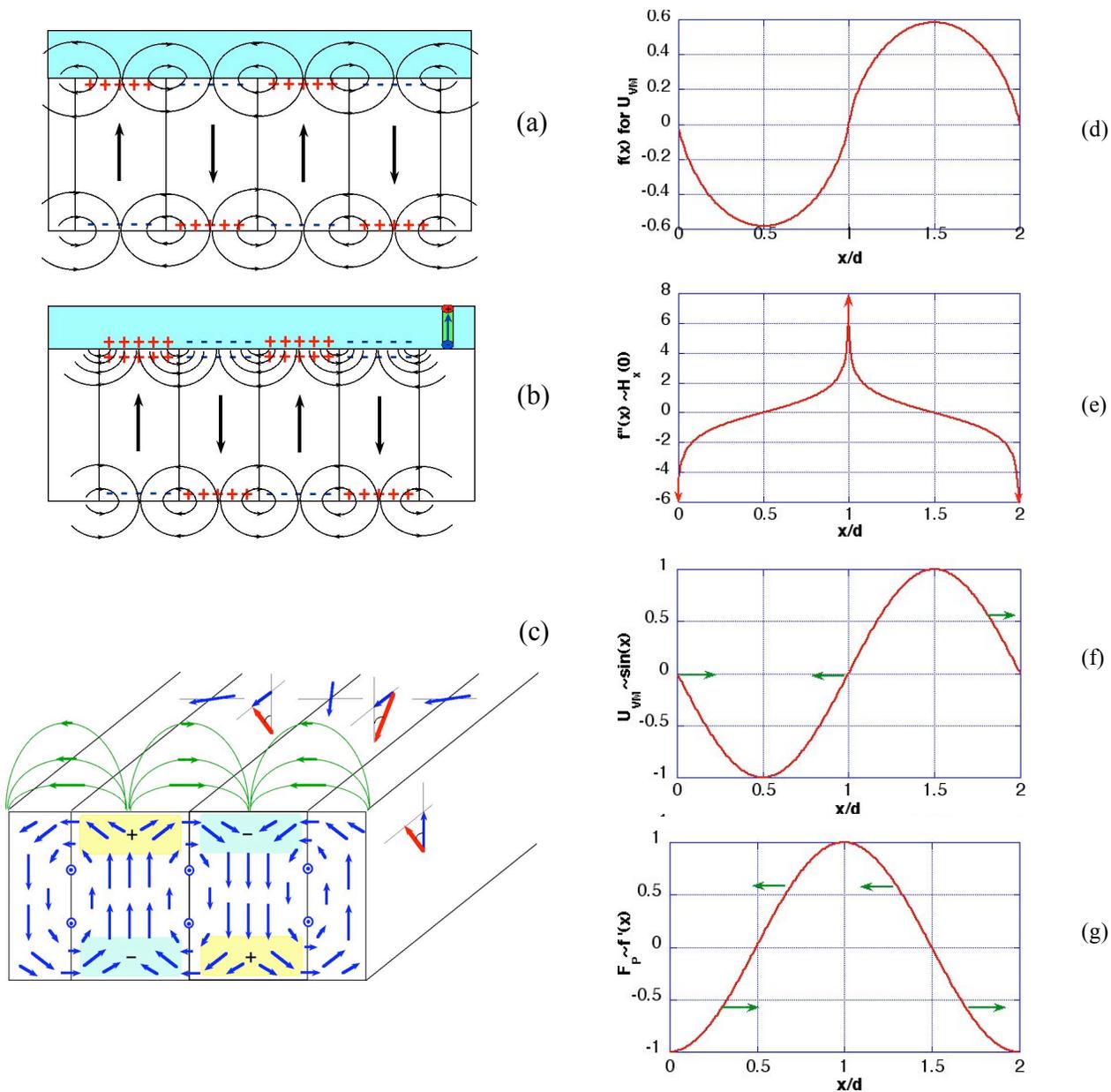

Fig.6 (a) Schematics of the domains and stray fields in high-Q FM films with perpendicular anisotropy. (b) At $T<T_c$, the magnetic stray fields from the FM domains are expelled from the top SC layer and doubled inside the FM resulting in the reduction of the domain width (see text). SC vortex interacts with surface magnetic charges in the domains as a magnetic monopole. The interaction potential and the force (it diverges for zero width domain walls) calculated following [35] are shown in (d) and (e). In low-Q Py, the magnetization distribution within domains is inhomogeneous (c) although the stray field pattern is qualitatively similar to (a). We approximate the vortex-magnetization interactions with a sinusoidal function (f) yielding the appropriate force landscape (g). Arrows in (f-g) show the force direction.